\documentclass[runningheads,a4paper]{llncs}

\usepackage{appendix}
\usepackage{adjustbox}
\usepackage{amssymb}
\usepackage{graphicx}
\usepackage{multirow}
\usepackage{cite}
\usepackage{float}
\usepackage{threeparttable}
\usepackage{xspace}
\usepackage{tabularx}

\newcommand{\ie}{{\it i.e.},\ }
\newcommand{\eg}{{\it e.g.},\ }

\usepackage{xcolor}

\begin{document}

\title{Android Malware Clustering through \\ Malicious Payload Mining}

\author{}
\institute{}

\author{Yuping Li\inst{1}\and
  Jiyong Jang\inst{2}\and
  Xin Hu\inst{3}\and
  Xinming Ou\inst{1}}
\institute{
  University of South Florida\\
  \email{yupingli@mail.usf.edu, xou@usf.edu}\and
  IBM Research\\
  \email{jjang@us.ibm.com}\and
  Pinterest\\
  \email{huxinsmail@gmail.com}
}

\maketitle

\begin{abstract}

Clustering has been well studied for desktop malware analysis as an effective
triage method. Conventional similarity-based clustering techniques,
however, cannot be immediately applied to Android malware analysis due to
the excessive use of third-party libraries in Android application development
and the widespread use of %
repackaging in malware development.
We design and implement an Android malware clustering system through
iterative mining of malicious payload and checking whether malware samples
share the same version of malicious payload.
Our system utilizes a hierarchical clustering technique and an efficient
bit-vector format to represent Android apps.
Experimental results demonstrate that our clustering approach achieves
precision of 0.90 and recall of 0.75 for Android Genome malware dataset, and average 
precision of 0.98 and recall of 0.96 with respect to manually verified ground-truth.

\end{abstract}

\section{Introduction}
\label{sec:introduction}

Triaging is an important step in malware analysis given the large number of
samples received daily by security companies. Clustering, or grouping malware
based on behavioral profiles is a widely-studied technique that allows analysts
to focus their efforts on new types of malware.
Multiple static~\cite{jang2011bitshred,Ye:2010jn}, %
dynamic~\cite{bayer2009scalable,Rieck:2011kw}, and hybrid~\cite{Hu:2013bg %
} analysis based clustering techniques have been proposed in the desktop malware domain.

With the rapid growth of Android smart devices, malicious Android apps
have become a persistent problem.
Security companies receive a list of (potential zero-day) malware on a daily basis~\cite{report}.
Those apps that present certain suspicious behaviors but are not detected by any
existing anti-virus scanners need to be further analyzed manually.
Conducting clustering on those incoming malware apps can allow the analysts to
triage their tasks by (a) quickly identifying
malware that shares similar behaviors with known existing malware so they may
not allocate much resources on it; and (b) selecting a few
representative apps from each new malware cluster to prioritize their analysis.

We often observe that existing approaches to group Android malware based on
their behaviors have provided limited capabilities.
For example, existing Android malware detection products may report a family
name for a detected sample;
however, samples from one family can have multiple different versions of
malicious code segments
presenting significantly different behaviors. 
Therefore, the malware family information provided by AV products can be
incomplete to describe crucial malicious code segments of Android malware.

Existing overall similarity analysis based clustering system cannot
be immediately applied for Android malware clustering because the malicious
code segments often constitute only a small fraction of an Android malware
sample.
In desktop malware clustering, the static or dynamic features are first
extracted from target samples. Then a clustering algorithm (\eg hierarchical
agglomerative clustering) is applied to group the samples such that samples
within the same resulting group share high level of overall similarity.
However, we note that overall similarity analysis performs poorly in Android
malware clustering because of two common practices in Android malware development.

The first practice is repackaging. Malware writers may embed the malicious
code inside an otherwise legitimate app, in which case the real malicious
code segment is likely to be small compared to the original benign app.
Our analysis shows that the ratio of the core malicious code segments to the
entire app for
a collection of 19,725 malware samples is between 0.1\% and 58.2\%.
Given the small percentage of malicious code segments,
the conventional clustering approach that is based on overall code similarity
will not work well.
For example, two malicious samples from different families can
be repackaged based on the same original benign app, thus presenting high level
of overall similarity. Likewise, Android malware
variants with the same malicious code of one family can be repackaged into
different original benign apps, thus presenting low level of overall similarity.

Another practice is utilizing shared library code.
Android apps often include a variety of third-party libraries to
implement extra functionalities in a cost-effective way.
If the library code size is too large compared
to the rest of the app, samples from different
malware families may be clustered together simply
because they share the same libraries.
We measured the library code proportion of the 19,725 malware samples.
For 13,233 of the samples that used at least one legitimate library,
we found that the average library code ratio is 53.1\% in terms of number of
byte code instructions.
This means a large portion of an Android app belongs to libraries.
One approach to prevent those libraries from ``diluting'' the
malicious code segments is to use a whitelist~\cite{grace2012riskranker,
  crussell2012attack, egele2013empirical, chen2014achieving, chen2015finding} to
exclude all library code.
However, previous work leverages only the names of libraries while building a
whitelist as opposed to the content of libraries.
We observed that malware authors injected their malicious code under
popular
library names, such as {\tt com.google.ssearch}, {\tt com.android.appupdate},
{\tt android.ad.appoffer}, and {\tt com.umeng.adutils}.
Consequently, na\"ive whitelisting approaches inadvertently remove certain
malicious payloads together with the legitimate library code from analysis.
We found that about 30\% of our analyzed Android malware families
disguise their malicious payload under popular library names.

Due to the above two reasons, directly applying overall similarity analysis on
Android apps will not be effective for Android malware analysis.
A major challenge is to precisely identify the malicious code segments
of Android malware.
For simplicity, we refer to the core malicious code segments of Android malware
as {\it malicious payload}. A payload can be an added/modified part
of a repackaged malware app, or the entire code of ``standalone'' malware app
excluding legitimate library code.

In this paper we propose an Android malware clustering approach
through iterative mining of malicious payloads.
Our main contributions include:

\begin{enumerate}

\item We design and implement an Android malware clustering solution through checking if apps
    share the same version of the malicious payloads.
    By reconstructing the original malicious payloads, %
    our approach offers an effective
    Android malware app clustering solution along with fundamental insights into
    malware grouping.

\item We design a novel method to precisely remove legitimate library code from
    Android apps,
    and still
    preserve the malicious payloads even if they are injected under popular
    library names.

\item We conduct extensive experiments to evaluate the consistency and
      robustness of our
      clustering solution.
    Our experimental results demonstrate that our clustering approach achieves
    precision of 0.90 and recall of 0.75 for Android Genome malware dataset, and
    average precision of 0.984 and recall of 0.959 regarding manually verified
    ground-truth.

\end{enumerate}

\section{Overview of Android Malware Clustering System}
\label{sec:clustering}

Rather than directly conducting overall similarity analysis between Android
malware samples, we first design a solution to precisely remove legitimate
library code from Android apps.
We consider the shared code segments (excluding legitimate library code) between the analyzed Android apps 
as candidate payload, and find all of the input Android apps through 
pairwise intersection analysis.
For a group of $n$ apps, each input app will contribute to $n-1$ versions of
candidate payloads.

After extracting all candidate payloads, we conduct traditional clustering 
analysis on all candidate payloads to group similar ones together.
Base on several key insights that are learned from analyzing candidate payload clustering results,
we design an effective approach to iteratively mine the payload clusters that are most 
likely to be malicious, and make sure that each input app
will only contribute one version of malicious payload.
Finally, we use the identified malicious payload clusters and payload-to-app 
association information to group the input Android malware apps.
We describe this process in more details below.

\begin{figure*}[ht]
  \centering
  \includegraphics[width=\columnwidth]{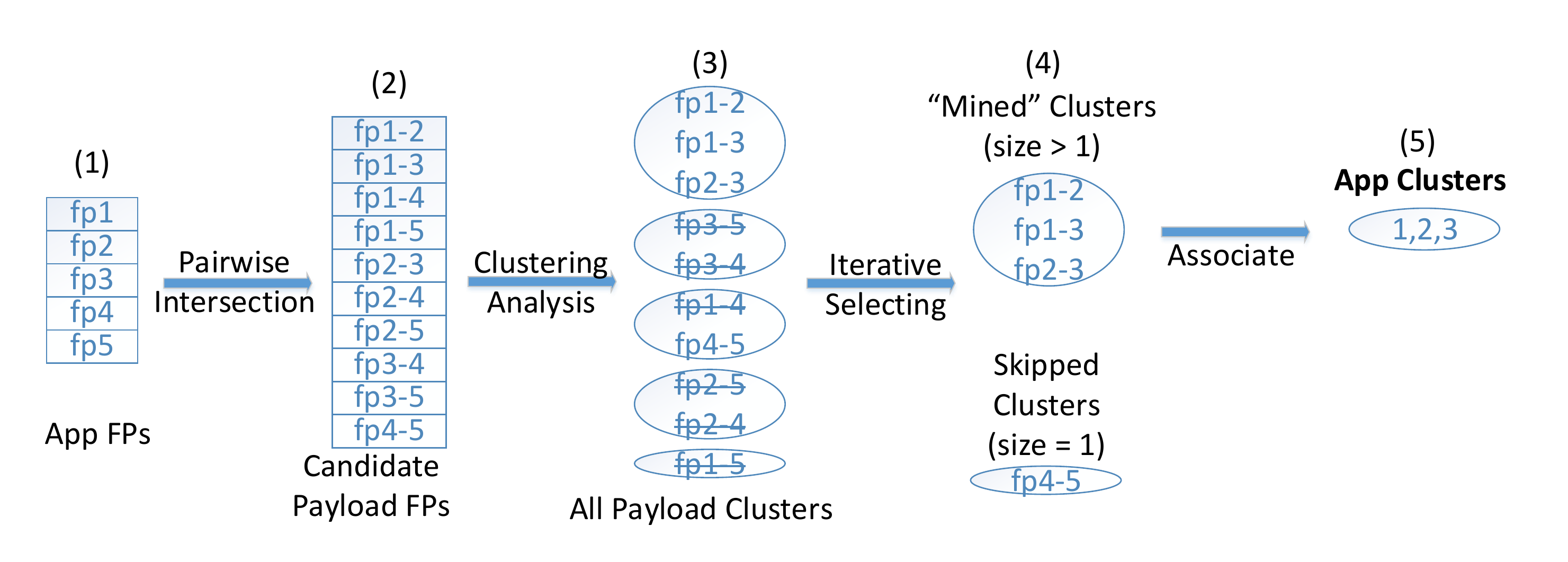}
  \caption{Overview of the clustering system with five Android malware samples.}
  \label{fig:comm_multi}
\end{figure*}

Figure~\ref{fig:comm_multi} illustrates the overview of the clustering analysis system with five
malware samples.

\begin{enumerate}
  \item {\bf Library code removal:} We convert malware samples into fingerprint representation,
    and design an effective approach to precisely remove legitimate library code
    from each app fingerprint.
    We denote the library-removed app fingerprints as {\sf fp1}, {\sf fp2}, {\sf fp3}, {\sf fp4}, and {\sf fp5} accordingly.

  \item {\bf Candidate payloads extraction:} We conduct a pairwise intersection 
    analysis to extract all shared code segments (\eg candidate payloads) between input apps.
    Relying on the app fingerprint representation, we create 
    candidate payload fingerprints, and record the payload-to-app 
    association information. For example, {\sf fp1-2} indicates that this
    candidate payload is extracted from malware sample 1 and 2.

  \item {\bf Candidate payloads clustering:} We then perform hierarchical clustering
    on all candidate payloads with a predefined
    clustering similarity threshold ${\it \theta}$, \eg the candidate
    payload fingerprints {\sf fp1-2}, {\sf fp1-3}, and {\sf fp2-3} are grouped
    together as the largest payload cluster based on the overall payload similarity.

  \item {\bf Malicious payload mining:} 
    After removing legitimate libraries, similar malicious payloads 
    extracted from samples in the same malware family will
    become more {\it popular}\footnote{Further intuition explanation and popularity criteria are included in
      Section~\ref{sec:mining}.} due to the ``legitimate'' reason of code reuse.
    Therefore, we design an iterative approach to mine the popular 
    payload clusters from the clustering results, which are more likely malicious payload.
    For instance, candidate payload cluster containing {\sf fp1-2}, {\sf fp1-3}, and {\sf fp2-3}
    is selected as the most popular cluster.
    To ensure that each input app only contributes one version of final malicious payload,
    we simultaneously update the remaining payload clusters.
    \eg fingerprints {\sf fp1-4}, {\sf fp1-5}, {\sf fp2-4}, {\sf fp2-5}, {\sf fp3-4}, and
    {\sf fp3-5} are then skipped because malware sample 1, 2 and 3 have already been ``used''.

  \item {\bf Malicious samples grouping:} We group the original Android samples
      based on payload mining results and payload-to-app association information 
      such that the samples within each app cluster contains the same version of the malicious payload.
      For example, malware samples 1, 2, and 3 are grouped together
      based on the selected candidate payload cluster containing {\sf fp1-2}, {\sf fp1-3}, and {\sf fp2-3}.

\end{enumerate}

\section{App Fingerprint Representation and Utilization}
\label{sec:background}

As we can see from Section~\ref{sec:clustering}, the clustering system
requires an effective fingerprint representation to denote input Android apps and candidate payloads.
Ideally, the fingerprint needs to be constructed from
the code segments of the input app
and support two fundamental operations:
precisely removing legitimate code, correctly extracting shared app code.

Based on these requirements, we decide to represent Android apps 
as bit-vector fingerprints, by encoding the features that are extracted from app code through feature hashing~\cite{jang2011bitshred,santos2011semi,hu2013mutantx}.
The value of each bit in the generated fingerprint is either 0 or 1,
indicating whether the corresponding app has a specific feature or not.

This bit-vector format enables us to precisely remove legitimate
library code (Section~\ref{sec:library}), extract shared code segments 
(Section~\ref{sec:extraction}), and reconstruct the original 
malicious payload (Section~\ref{sec:reconstruct}) by utilizing the bit manipulation capability.

\subsection{Fingerprint Generation and Fingerprint Comparison}
\label{sec:fp_gen}

\begin{figure*}[ht]
  \centering
  \includegraphics[width=\columnwidth]{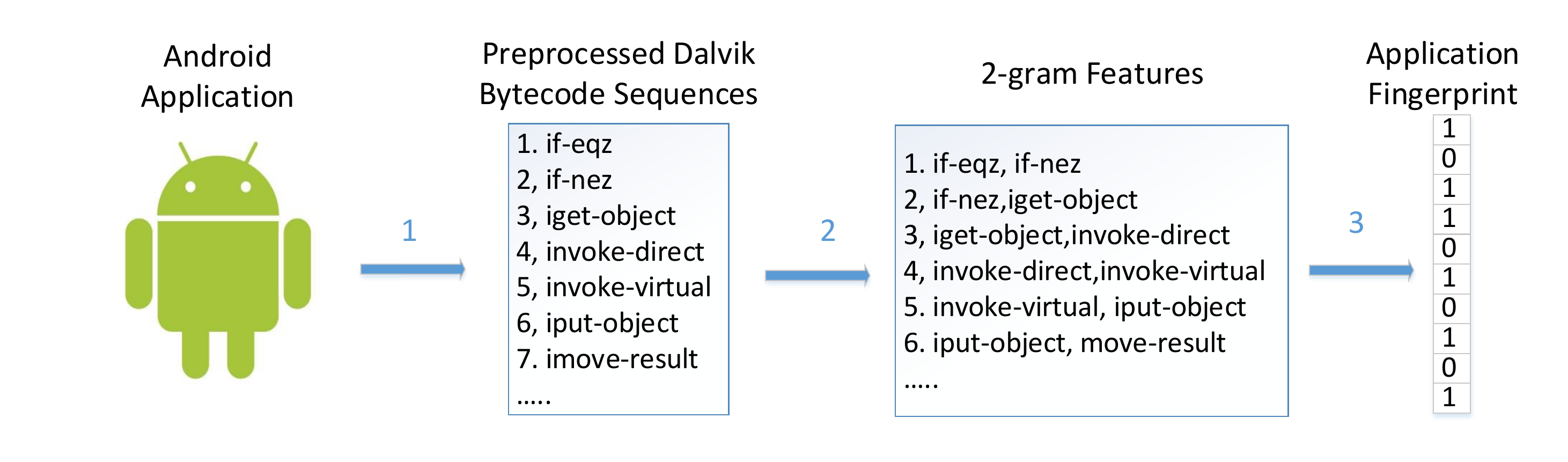}
  \caption{Overall fingerprint generation procedure}
  \label{fig:fpgen}
\end{figure*}

In this work, we use {\it n}-gram sequence of Dalvik bytecode to denote
an Android app feature, and use a {\it bit-vector} fingerprint to represent the extracted features.
The overall fingerprint generation process is shown in Figure~\ref{fig:fpgen}.

For each Android app, we first use Dexdump~\cite{dexdump} to disassemble {\tt classes.dex} into Dalvik bytecode, then preprocess the Dalvik bytecode sequences to only include the major distinctive information
and extract the {\it n}-gram features from the preprecessed bytecode sequences.
We follow similar approach to extract the distinctive information (\eg bytecode opcode) for feature construction as Juxtapp~\cite{hanna2013juxtapp}.
Since feature space is vital to support the key operations designed in this work, we decide to increase the feature space by including more generic but meaningful information from each bytecode instruction.
The major distinctive information is separated into 4 categories and summarized in Table~\ref{table:features}.
Besides the feature differences shown in Table~\ref{table:features}, we extract the {\it n}-gram features at the function level, while Juxtapp extracts {\it n}-gram features at the basic block level. 
For simplicity, we only show the Dalvik bytecode opcode sequences as the distinctive instruction information in Figure~\ref{fig:fpgen}.

\begin{table}[h]
\caption{Major feature categories and main differences comparing with Juxtapp}
\label{table:features}
\centering
\scalebox{0.97}{
\begin{tabular}{|l||c|c|c|}
\hline
Feature Category & Examples & {Our Approach} & {Juxtapp} \\
\hline
\hline
Dalvik bytecode opcode sequences & sget, goto, return & \checkmark & \checkmark \\
\hline
Java VM type signatures & Z({\it Boolean}), B({\it byte}) & \checkmark & \\
\hline
String value of {\tt const-string} instructions & - & \checkmark & \\
\hline
Type signatures for ``invoked'' functions & f(I,[B)Z & \checkmark & \\
\hline
\end{tabular}
}
\end{table}

After extracting all the $n$-gram features, we then encode all the features in 
a bit-vector format fingerprint through {\bf feature hashing} technique using {\tt djb2} hash function.
During feature hashing process, we use a tuple $A(i, j)$ to represent a feature position, in which {\it i} is the function offset indicating from which function the particular {\it n}-gram feature is extracted, and {\it j} is the bytecode offset indicating the position of the {\it n}-gram feature within the corresponding function.
Then the feature-to-bit information is stored in a map, in which the key is the bit index within the fingerprint indicating where the feature is stored, and the value is the list of feature tuples that are 
mapped to the bit location. 
With increased feature space, we hope to reduce majority ofthe random feature collisions, and allow each bit index to represent the same {\it n}-gram feature content.

Similar to the complete Android apps, individual legitimate libraries
and the candidate malicious payloads are also represented in the same size of bit-vector fingerprints.
The concrete $n$-gram size and the fingerprint size used for clustering are determined through
analyzing the collision rate of random features, which is discussed in Section~\ref{subsec:collision}.

To measure the similarity between two fingerprints, we use the Jaccard index, or
the Jaccard similarity, which is defined as the size of intersection divided by the
size of union of two sets.
Since each fingerprint is a bit-vector, we leverage cache-efficient bit-wise AND
($\wedge$) and bit-wise OR ($\vee$) operations to compute the intersection and
the union. Then, the similarity of two fingerprints ${\sf fp_a}$ and ${\sf
  fp_b}$ is defined as follows:

\begin{equation}\label{eq:sim_func}
Similarity({\sf fp_a}, {\sf fp_b}) = {{S({\sf fp_a} \wedge {\sf fp_b})} \over {S({\sf fp_a} \vee {\sf fp_b})}},
\end{equation}
where $S(\cdot)$ denotes the number of 1-bits in the input.

Our fixed-sized bit-vector fingerprint representation also allows us to
easily measure containment ratio in a similar fashion:

\begin{equation}\label{eq:con_func}
Containment({\sf fp_a}, {\sf fp_b}) = {{S({\sf fp_a} \wedge {\sf fp_b})} \over {S({\sf fp_a})}},
\end{equation}
which measures how much of the content of ${\sf fp_a}$ is contained in ${\sf fp_b}$.

\subsection{Fingerprint based Library Code Removal}
\label{sec:library}

\begin{figure}[!h]
\begin{minipage}[t]{0.5\linewidth}
  \centering
  \includegraphics[width=\columnwidth,height=2.5cm]{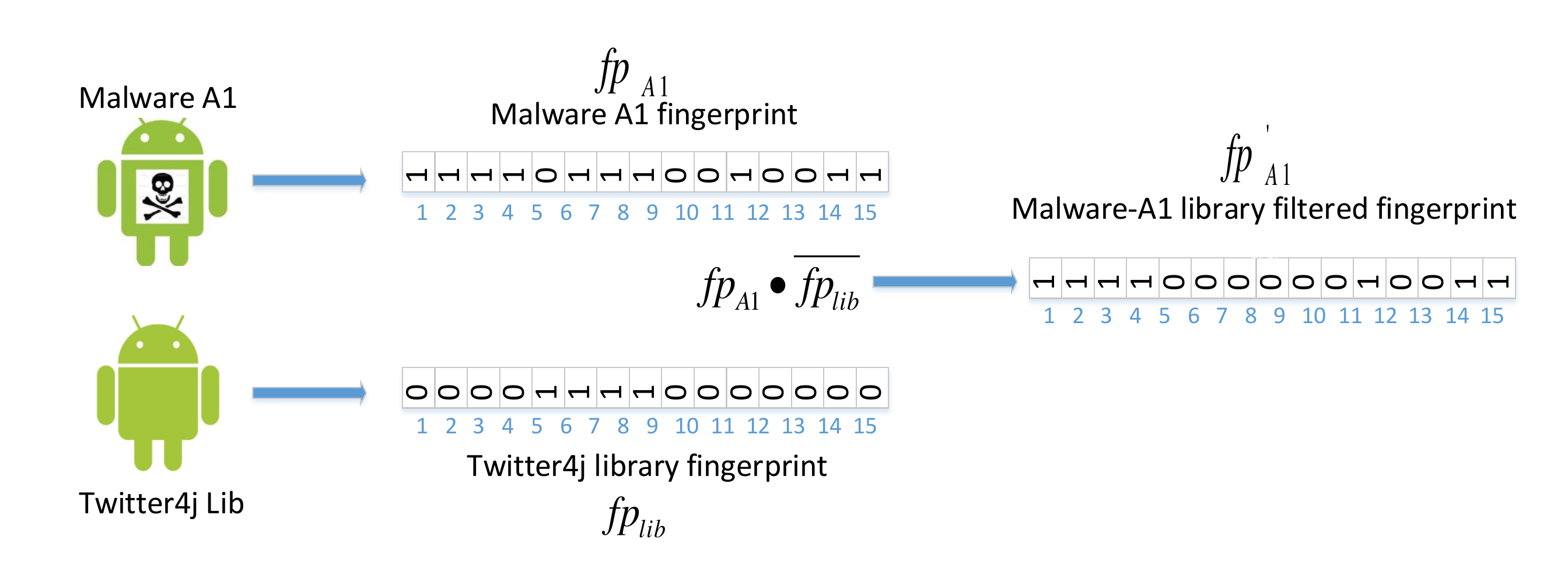}
  \centering
  \caption{Example procedure to safely remove legitimate ``twitter4j'' library code}
  \label{fig:library}
\end{minipage}
\hspace{0.1cm}
\begin{minipage}[t]{0.5\linewidth}
  \centering
  \includegraphics[width=\columnwidth, height=2.5cm]{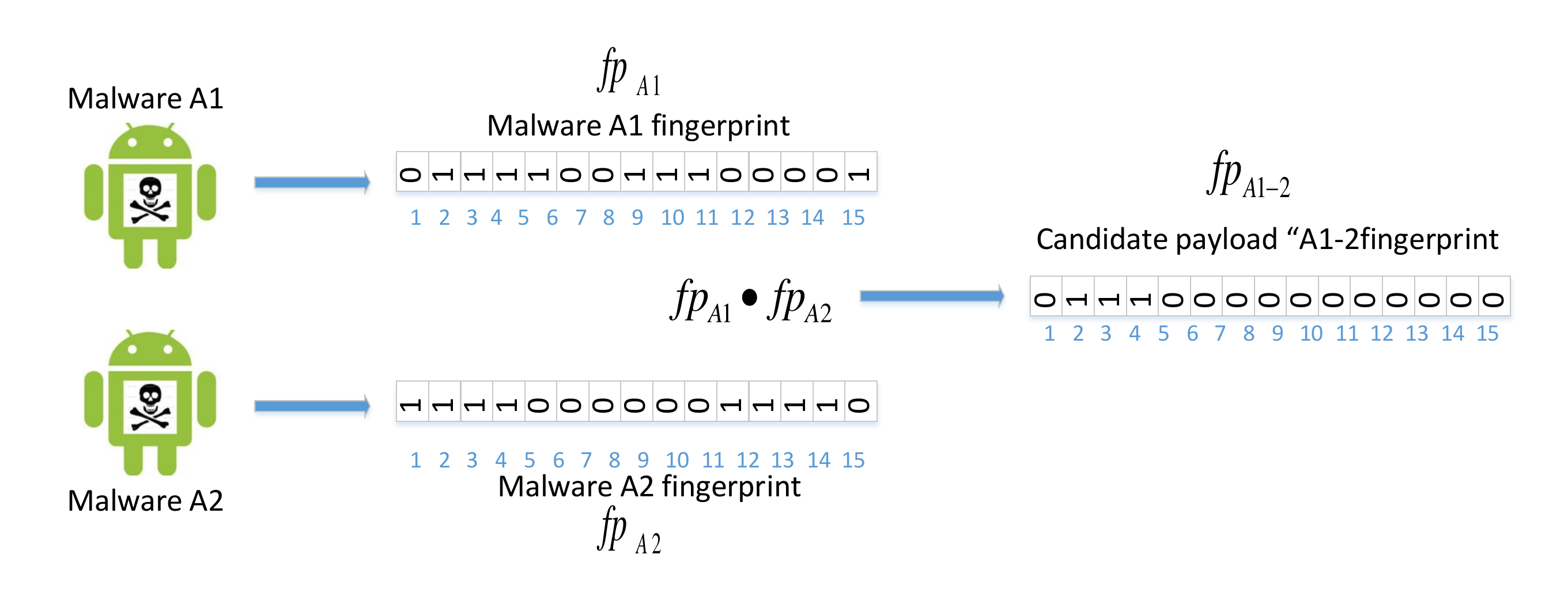}
  \caption{Extracting a candidate payload from two malware applications}
  \label{fig:extraction}
\end{minipage}
\end{figure}

To precisely remove legitimate library code without excluding a possibly
injected
malicious payload, we exclude legitimate library code from an app by
removing the {\it library-mapped} bits from the app bit-vector fingerprint.
For each legitimate library, we collect its official jar file and disassemble
it into Dalvik bytecode sequences; then apply the same feature hashing technique
to map the \textit{n}-gram features of the library code into a bit-vector
fingerprint ${\sf fp_{lib}}$.
We then flip all the bits in the library fingerprint to get $\overline{\sf fp_{lib}}$.
Since the same features contained in an Android app and the library
are mapped to the same bit positions in their fingerprint representation,
we can exclude library-mapped bits from an app fingerprint by bit-wise ANDing
$\overline{\sf fp_{lib}}$ and ${\sf fp_{app}}$.
Figure~\ref{fig:library} demonstrates the overall procedure to safely remove
legitimate {\tt twitter4j} library code from a malware sample.

We first conduct statistical analysis for the disassembled apps to
identify the embedded legitimate libraries, and record the years when the target
samples were created.
We then obtain\footnote{We randomly select one version of library in each
  year in case there are multiple versions of libraries released within the same
  year.} the officially released library jar
files to create the corresponding library fingerprints, and remove the library code
from the analyzed apps.
The library code removal process is applied only when an app contains code
snippets that are defined under corresponding library namespaces.

In our implementation, each library is represented with an individual fingerprint.
We encode multiple versions of the same library together in a single library
fingerprint.
This aggregated library representation may cause potential feature collision
between the app code and the irrelevant versions of the library code.
However, we empirically demonstrate in Section~\ref{subsec:lib_acc} that
the library code removal process is precise %
because different versions of the same library typically share high level
of code similarity due to code reuse, and the size of the single library is
often smaller than the entire app.

\subsection{Fingerprint based Candidate Payload Extraction}
\label{sec:extraction}

The next operation is to extract malicious payloads from malware
samples.
We consider the shared code segments (after excluding legitimate libraries) 
between each malware sample pair to be a candidate malicious payload.
For a group of malware samples, we obtain the intersection of every fingerprint
pair of library-excluded samples, and consider the shared 1-bits between the
sample fingerprints as a candidate payload fingerprint.

Figure~\ref{fig:extraction} describes the intersection analysis procedure to
extract a candidate malicious payload at a high level.
For two malware samples we first build their fingerprints and exclude the 
legitimate library bits from the fingerprints. Then we pinpoint their shared 1-bits (\eg
bits index 2, 3, and 4) as potentially malicious\footnote{malicious payload mapped} bits
and construct a candidate payload fingerprint.

During the candidate payload extraction process, we keep track of the
association information between the candidate payload (\eg {\tt A1-2}) and the
corresponding samples (\eg {\tt A1} and {\tt A2}).
We subsequently use the payload-to-app association information and the malicious payload mining results to group malware samples.

\subsection{Fingerprint based Malicious Payload Reconstruction}
\label{sec:reconstruct}

Using the bit-vector fingerprint representation, we can also  
define the cluster fingerprint for a version of the candidate payload cluster
as the intersection of all the candidate payload fingerprints in the cluster.
The 1-bits contained in the resulting cluster fingerprint 
can be viewed as the shared malicious bits for all input
apps that share the same version of malicious payload.

Using the identified malicious bits from app fingerprints,
we can then reconstruct the corresponding malicious payload code by checking the feature-to-bit mapping information that was recorded during feature hashing, which can be viewed as the reverse procedure of fingerprint generation.
Given the identified malicious bits, we locate the feature tuples that are mapped to those identified malicious bits.
We use each retrieved feature tuple to locate the \textit{n} 
lines of code where the \textit{n}-gram feature is extracted, then reconstruct complete malicious code sequences by
properly stitching the identified \textit{n} lines of code segments together.

In practice, feature collision is possible but becomes negligible with appropriate {\it n}-gram size and fingerprint size, thus we will rarely recover the irrelevant code.
To certain extent, payload code reconstruction compensates feature hashing
collisions (\eg resulting in missing \textit{n}-grams) as far as the missing
$n$-gram is within the overlapped original code sequences of recovered features.
The reconstructed malicious payload code can be further inspected to verify its maliciousness.

\section{Malicious Payload Mining}
\label{sec:mining}

{\bf Key insights:} (a) In practice, when feature hashing is configured to
have a low collision rate,
malware app fingerprints will not contain a large number of shared 1-bits
unless they do share certain common features (\eg payload code snippets).
(b) Likewise, if a target dataset contains malware samples that do share the
same version of the malicious payload, then the candidate payload fingerprints
extracted from those samples will contain similar shared 1-bits and
be automatically clustered into the same group.
(c) After removing legitimate library code from an app, similar malicious
payloads have higher chances to form a {\it larger cluster}
than the ones related to less popular libraries or coincidentally shared code
segments.
(d) Compared to coincidentally shared code segments, similar malicious payloads will
have a {\it larger shared code base} because of ``legitimate'' reason of code
reuse in the same malware family, and the fingerprints for the malicious
payloads will have a larger amount of shared 1-bits.

Based on the above key insights, we design the following strategies to
iteratively
select representative candidate payload clusters based on payload {\it
  popularity}, which is determined based on the three criteria:
the entry size of a payload cluster {\it l},
the number of distinct apps associated with a payload cluster {\it m}, and
1-bits count of a payload cluster fingerprint {\it k}.

\begin{itemize}
\item{We count the number of candidate payload fingerprint entries in each
    cluster, and maximize the possibility of extracting core malicious payloads
    by selecting the clusters with the largest number of payload fingerprint
    entries.
    Payload cluster size {\it l} is a direct indicator for the popularity of the
    shared code segments between malware samples, and such popular shared code
    is a good candidate for one version of malicious payloads since we have
    already filtered out popular legitimate library code.
}
\item{We measure the distinct apps {\it m} that contribute to generating
    candidate payload fingerprints of each cluster, and select the clusters with
    the largest number of distinct apps if they have the same number of payload
    entries.
    Payload clusters that contain a large number of unique payload entries are
    often associated with a large number of distinct apps, and we use this app
    association information to break the tie in case the number of cluster
    entries are the same since distinct apps can be considered as another sign
    of comparative popularity.
}
\item{We obtain the intersection bits {\it k} of payload fingerprint entries in each cluster as the cluster fingerprint. If two clusters are associated with the same number of distinct apps, we then select the one with the larger number of 1-bits in its cluster fingerprint.
    In this way, we can extract the payload with a larger code size, and it
    helps to increase the likelihood of getting malicious payloads together with
    shared libraries, and we subsequently exclude possibly remaining libraries
    later.
}
\item{During cluster selection, we keep track of which apps have been used to
    generate candidate payload fingerprints in the previously selected clusters,
    and consider already-selected apps as ``inactive''.
    We update the remaining payload clusters by removing candidate fingerprint entries that are associated with ``inactive'' apps.
    Skipping such fingerprints allows us to extract one version of the malicious payload from each app.
    This helps to merge all the shared core malicious code together, 
    and only extract the widely shared malicious code between all apps,
    which also helps to reduce the probability of extracting 
    non-malicious payload code.
}
\item{
    We omit a payload cluster if the corresponding cluster fingerprint contains
    less than
    the minimum {\it k} number of 1-bits, meaning that the extracted code
    segments are too small.
    It forces the algorithm to break the current large payload cluster into
    smaller clusters with a larger code size, and prevent different malware
    families from being clustered together.
    We set the minimum
    number of 1-bits {\it k} to 70 since the majority of the analyzed Android
    malware app fingerprints had more than 70 1-bits.
}
\item{
    We exclude a candidate payload cluster if it becomes empty after the update
    in the last step, or if the number of payload fingerprint entries is too
    small (\eg ${\it l} = 1$).
    This is because
    Clusters with only a single candidate payload entry provide little
    additional popularity information, and are more likely to contain less
    popular libraries or other coincidentally shared code snippets. We consider
    malware samples associated with such payload clusters as unclustered, and
    the unclustered app is evaluated as a singleton.
}
\end{itemize}

The shared payloads between Android samples can be library code segments,
malicious
payloads, copy-and-pasted code segments, or other coincidentally shared code
segments.
The above payload mining strategy enables us to select the most likely malicious
candidate payload groups.
Legitimate non-library reused code may be collected together with malicious payload only if it is shared across a significant number of apps. 
Otherwise, the less popular legitimate non-library code will be evidently excluded during the (popularity-based) payload mining procedure.
If the same benign app is indeed used by many malware apps, we can further exclude original benign app code (\ie the legitimate non-library reused code) in a similar way to remove library code using a benign app fingerprint.

\section{Optimize Overall Clustering Efficiency}
\label{sec:optimization}

According to the previously discussed malicious payload mining procedure, we will generate $\frac{n \times
 (n-1)}{2}$ versions of candidate payload fingerprints given \textit{n} malware samples, but the
hierarchical clustering algorithm also has a quadratic complexity with respect to
the number of analyzing targets.
  Due to the overall quartic complexity of the algorithm, directly using it to analyze large number of samples becomes a time-consuming task.
  Therefore, we further develop two methods to improve the scalability of the
  clustering analysis procedure, and hereafter refer them as {\sf
    Opt-1}, and {\sf Opt-2}. %

      \subsection{{\sf Opt-1}: Optimize Each Pairwise Computation}

      The first method to speed up the overall clustering process
      is to optimize each pairwise computation.
      Broder proposed minHash~\cite{broder1997resemblance} to quickly estimate the Jaccard
      similarity of two sets without explicitly computing the intersection and the
      union of two sets.
      By considering our bit-vector fingerprint as a set, we apply minHash to
      further transform a large fingerprint into a smaller size signature, and
      calculate the similarity of minHash signatures to estimate the Jaccard
      similarity of the original fingerprints.

      To apply minHash, we define a minHash function output of our bit-vector
      fingerprint $h({\sf fp})$ to be the first non-zero bit index on a randomly permutated
      bits order of the fingerprint.
      We then apply the same minHash function to two fingerprint ${\sf fp_a}$ and ${\sf fp_b}$.
      This will generate the same minHash value when the first non-zero bit indexes
      for two fingerprints ${\sf fp_a}$ and ${\sf fp_b}$ are the same.
      Since the probability that the firstly encountered bit is a non-zero bit for
      ${\sf fp_a}$ and ${\sf fp_b}$ is conceptually the same as Jaccard similarity $Similarity({\sf fp_a},
      {\sf fp_b})$~\cite{leskovec2014mining}, we use such probability $Pr[h({\sf fp_a})=h({\sf fp_b})$
      to estimate the original Jaccard similarity.

      The probability estimation becomes more accurate if more independent minHash
      functions are used together.
      Formally, we define a minHash signature $sig({\sf fp})$ to be a set of \textit{k}
      minHash function values extracted from \textit{k} round of random permutations
      over the fingerprint, and represent it as follows:
      $sig({\sf fp})=[h_1({\sf fp}), h_2({\sf fp}), ..., h_k({\sf fp})]$.
      We denote the similarity of two minHash signatures as the ratio of equal
      elements between $sig({\sf fp_a})$ and $sig({\sf fp_b})$.

      Instead of maintaining \textit{k} random permutations over the bit-vector, we
      follow a common practice for using minHash technique and use \textit{k}
      different hash functions to simulate \textit{k} random permutations, where each
      hash function maps a bit index to a value.
      In order to create \textit{k} hash functions, we first generate \textit{k}
      random numbers, then use FNV~\cite{fnv} hash algorithm to produce a basic hash
      output for each bit index, and finally apply XOR operation between each random
      number and the hash output to get the \textit{k} hash outputs.
      For each hash function, we select the smallest hash value (to simulate the
      first non-zero bit index) over all of the
      bit indexes of the fingerprint as the final hash output.

      Note that the FNV hash value and the \textit{k} random numbers are all 32 bits
      unsigned integers, and they can be used to safely simulate random
      permutation over 512MB bit-vector fingerprint.
      In practice, the \textit{k} value usually needs to be larger than 100 to
      generate good enough results~\cite{leskovec2014mining}.
      We set \textit{k} to be 256 in our experiments, and thus convert each
      bit-vector fingerprint into a 1KB minHash signature.

      In order to evaluate the potential impact of {\sf Opt-1} on accuracy, we
      conduct two experiments on the smallest 50 malware families\footnote{We select
        those families since their maximum family size is
      under 100 and all the experiments for those families can be finished within 1 hour.}:
      one experiment ({\sf Exp-1}) with no optimization, and another experiment
      ({\sf Exp-2}) using {\sf Opt-1}.
      We used the clustering output from {\sf Exp-1} as a reference, and measured the
      precision and recall of the clustering output from {\sf Exp-2}.
      The precision and recall indicate how similar the two experiments results are,
      and are used to check the impact on accuracy brought by {\sf Opt-1}.

      Our experiments showed that on average {\sf Exp-2} took less than 83\% time to
      complete compared to {\sf Exp-1} for the analyzed families, and the average precision and
      recall of the clustering output were 0.993 and 0.986.
      {\sf Opt-1} significantly reduce the overall memory consumption with
      minHash signature representation and improve the pairwise computation efficiency
      with almost zero accuracy penalty.

      \subsection{{\sf Opt-2}: Employ approximate clustering}

      The previous optimization is still not sufficient for using the
      algorithm to analyze large scale malware samples.
      For instance, when analyzing with 2,000 samples, the algorithm will create
      1,999,000 candidate payloads, and it results in approximately $2.0 \times 10^{12}$
      pairwise comparison. Even 1\% of the total comparison
      still takes lots of computation resources.
      To resolve the scalability issue for a large dataset input, we further adopt
      prototype-based clustering technique~\cite{kim1996application, Rieck:2011kw} to achieve approximate clustering.

      Specifically, we randomly divide the target samples into small size (\eg 150) groups.
      For each group, we apply hierarchical clustering analysis on the shared payload
      within the group, and create a prototype fingerprint for each payload cluster by
      applying intersection analysis (to obtain all the shared 1-bit) among the
      payload fingerprints in each cluster. We then conduct hierarchical clustering
      analysis on all the collected prototype fingerprints.
      In this way, we represent a group of similar payload fingerprints with a single prototype
      fingerprint, and the algorithm proceeds with approximate clustering analysis
      using the prototype fingerprints instead of the original payload fingerprints.

      We design two experiments to evaluate the impact of {\sf Opt-2} on accuracy:
      one experiment ({\sf Exp-3}) using {\sf Opt-1} only, and another
      experiment ({\sf Exp-4}) using {\sf Opt-1} and {\sf Opt-2}.
      Due to the quartic complexity of the original algorithm, the overall
      analysis (using {\sf Opt-1} only) will get dramatically slower
      for analyzing larger number of malware samples.
      For instance, we found it takes about one day to analyze 1000 samples and
      more than five days to analyze 2000 samples for {\sf Exp-3}.
      In order to conduct the evaluation within reasonable amount of time,
      we randomly select 70\% of labeled samples from the largest 4 malware families
      and conduct the two experiments for each family.
      We used the clustering output generated by {\sf Exp-3} as reference, and
      measured the precision and recall of the clustering output generated by {\sf
      Exp-4} to evaluate the accuracy impact brought by {\sf Opt-2}.

      Our experiments showed that on average {\sf Exp-4} can speed up
      more than 95\% compared to {\sf Exp-3}, and the average
      precision and recall for the analyzed 4 families were 0.955 and 0.932.
      This optimization makes it feasible to apply our algorithm to
      analyze a bigger scale of malware families while providing a desirable
      trade-off option between speed and accuracy.

\section{Experiments}
\label{sec:experiments}

In this section, we describe the data preparation procedure, and report malware
clustering results and key findings of our experiments.

\subsection{Data Preparation}
We obtained a large collection of potentially malicious Android apps (ranging from late 2010 to early 2016) from
various sources, include Google Play, VirusShare~\cite{virusshare} and third party security companies.
In order to prepare ground-truth family labeling for the datasets, we queried the collected apps against
VirusTotal~\cite{virustotal} around April 2016, and used the scanning results to
filter out potentially ambiguous apps.

To assign family labels to the collected malware samples, we applied the
following steps:
(1) tokenized VirusTotal scanning results and normalized the contained keywords,
and then counted the total number of occurrences of each keyword.
(2) removed all the generic keywords such as {\tt Virus}, {\tt Trojan}, and {\tt
  Malicious}.
(3) detected keyword aliases by calculating the edit distances between keywords.
For example, {\tt Nickyspy},
{\tt Nickspy}, {\tt Nicky}, and {\tt Nickibot} were all consolidated into {\tt Nickispy}.
(4) assigned the dominant keyword as the family label for the sample.
A keyword was considered as dominant if it satisfied two conditions: (a) the
count of the keyword was larger than a
predefined threshold \textit{t} (\eg $t$=10), and (b) the count of the most
popular keyword was at least twice larger than the counts of any other keywords.

\begin{table}[ht]
\caption{Clearly Labeled Malware Families}
\label{table:family}
\centering
\scalebox{0.95}{
\begin{tabular}{|c|c||c|c||c|c||c|c||c|c|}
\hline
Name & Size & Name & Size & Name & Size & Name & Size & Name & Size\\
\hline
Dowgin & 3280 & Minimob & 145 & Erop & 48 & Vidro & 23 & Koomer & 15\\
Fakeinst & 3138 & Gumen & 145 & Andup & 48 & Winge & 19 & Vmvol & 13\\
Adwo & 2702 & Basebridge & 144 & Boxer & 44 & Penetho & 19 & Opfake & 13\\
Plankton & 1725 & Gingermaster & 122 & Ksapp & 39 & Mobiletx & 19 & Uuserv & 12\\
Wapsx & 1657 & Appquanta & 93 & Yzhc & 37 & Moavt & 19 & Svpeng & 12\\
Mecor & 1604 & Geinimi & 86 & Mtk & 35 & Tekwon & 18 & Steek & 12\\
Kuguo & 1167 & Mobidash & 83 & Adflex & 32 & Jsmshider & 18 & Spybubble & 12\\
Youmi & 790 & Kyview & 80 & Fakeplayer & 31 & Cova & 17 & Nickispy & 12\\
Droidkungfu & 561 & Pjapps & 75 & Adrd & 30 & Badao & 17 & Fakeangry & 12\\
Mseg & 245 & Bankun & 70 & Zitmo & 29 & Spambot & 16 & Utchi & 11\\
Boqx & 214 & Nandrobox & 65 & Viser & 26 & Fjcon & 16 & Lien & 11\\
Airpush & 183 & Clicker & 58 & Fakedoc & 26 & Faketimer & 16 & Ramnit & 9\\
Smskey & 166 & Golddream & 54 & Stealer & 25 & Bgserv & 16 &  & \\
Kmin & 158 & Androrat & 49 & Updtkiller & 24 & Mmarketpay & 15 &  & \\
\hline
\end{tabular}
}
\end{table}

Although our malware labeling process may look similar to
AVclass~\cite{sebastian2016avclass},
we developed the approach independently without the knowledge of the
AVclass; and both work was finished around the same time.
The unlabeled samples were not included in the malware dataset for clustering
analysis.
In summary, we collected 19,725 labeled malware samples from 68 different
families, and the detailed breakup of the malware samples is shown in
Table~\ref{table:family}.

Besides the above labeled malware dataset, we also collected Android Genome
malware
samples~\cite{zhou2012dissecting} to obtain an optimal clustering threshold, and
randomly selected a list of 10,000 benign samples from AndroZoo~\cite{allix2016androzoo} to evaluate the
accuracy of the library removal procedure.
In particular, we selected benign apps that were
created around the same time (before Jan 1st, 2016) as most of the labeled malware
samples, and their latest (Mar 2017) VirusTotal re-scanning results showed no
malicious labels.

\subsection{Feature Collision Analysis}
\label{subsec:collision}

\begin{figure}[!h]
\begin{minipage}[t]{0.5\linewidth}
  \centering
  \includegraphics[width=\columnwidth,height=4cm]{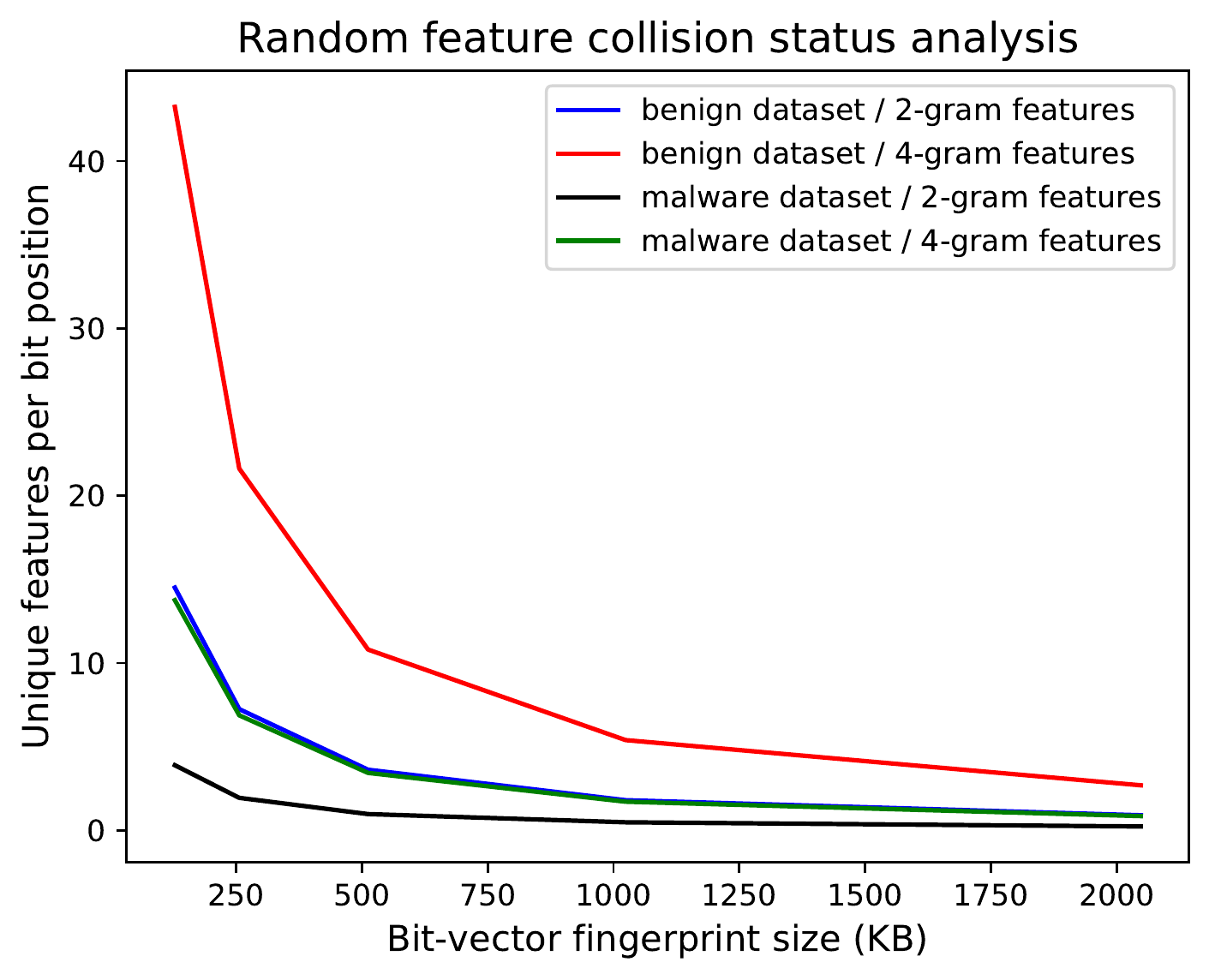}
  \centering
  \caption{Random feature collision status}
  \label{fig:collision}
\end{minipage}
\hspace{0.1cm}
\begin{minipage}[t]{0.5\linewidth}
  \centering
  \includegraphics[width=\columnwidth, height=4cm]{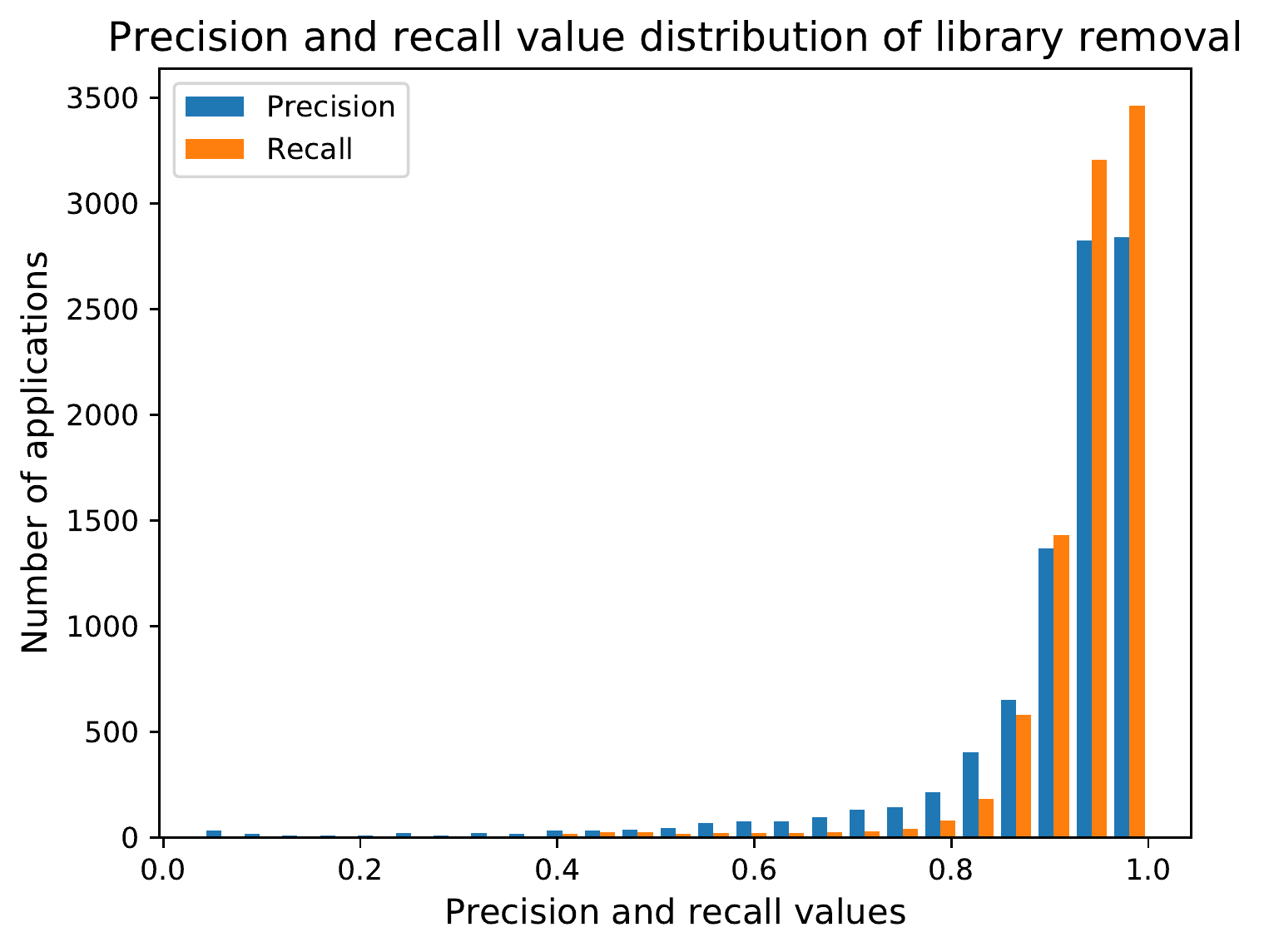}
  \caption{Benign apps lib removal accuracy}
  \label{fig:lib_acc}
\end{minipage}
\end{figure}

The accuracy of the proposed clustering system and the correctness of the reconstructed malicious payloads
relies on the assumption that unique features will be mapped
to unique bit locations within the bit-vector fingerprint.
Feature collision is directly impacted by two parameters: an $n$-gram size, and
a bit-vector fingerprint size.
To evaluate a feature collision rate, we varied the $n$-gram size (2 and 4) and
the bit-vector fingerprint size,
and then measured how many unique features were mapped to the same single bit
position, i.e., feature collision.
Figure~\ref{fig:collision} illustrates feature collision with regard
to different $n$-gram sizes and fingerprint sizes.

The graph shows that feature collision occurs more frequently when the
fingerprint size is small. The total number of unique features depends on the
$n$-gram size.
For the labeled malware, it was about
4.1 million
for 2-gram features, and 14.4 million for 4-gram features. 
And for the benign dataset, it was about 15.2 million
for 2-gram features, and 45.3 million for 4-gram features.
According to the {\it pigeonhole principle}, when putting {\tt N} unique
features into {\tt M} buckets,
with $N > M$, at least one bucket would contain more than one unique features.
This means that we need to set the bit-vector fingerprint size larger
than the total number of unique features to reduce feature collision.
Therefore, we set the default $n$-gram size to be 2 and default
fingerprint size to be 1024KB which provides 8,388,608 unique bit positions.
With the above configuration, the unique feature per bit value was reduced to
0.49 to process the labeled malware dataset.
Notice that the complete feature space is unlimited for our system due to the inclusion of
arbitrary string values, however the true unique features contained in a
certain dataset will be limited.

\subsection{Library Removal Accuracy}
\label{subsec:lib_acc}

Besides the random feature collision discussed in the previous section, 
it is also possible that feature collision may happen between the app code and the 
irrelevant versions of the library code.
To evaluate the library removal accuracy, we assumed the libraries used in
benign samples were
not purposefully manipulated, and measured the precision (\eg how much of the
removed code is true library code)
and recall (\eg how much of the true library code is removed) of library code
removal results
for the prepared benign samples.
Particularly, we considered the code that were defined under the official
library names in the benign samples
as ground truth library code, and created the {\it true library code}
  fingerprint ${\sf fp_{true}}$
by mapping all the features from the true library code to a bit-vector
fingerprint.
After removing the library code from each app, we identified the bit
positions that were presented in the original app fingerprint and were removed
subsequently; and used the
identified bit positions to generate {\it removed library code} fingerprint
${\sf fp_{removed}}$.
Using the containment ratio calculation function as discussed in
Section~\ref{sec:fp_gen}, library removal precision ${\sf P_{lib}}$ is defined as
${S({\sf fp_{true}} \wedge {\sf fp_{removed})}} \over {S({\sf fp_{removed}})}$,
and library removal recall ${\sf R_{lib}}$ is defined as $S{({\sf fp_{true}}
  \wedge {\sf fp_{removed}})} \over {S({\sf fp_{true}})}$,
where $S(\cdot)$ denotes the number of 1-bits in the bit-vector.

Figure~\ref{fig:lib_acc} depicts the library removal precision and recall for
the benign apps.
We observed that 9,215 benign apps contained at least one legitimate library,
and the median values for precision and recall were 0.94, 0.95, respectively.
We manually inspected certain corner cases with poor precision or recall.
The poor precision cases
were due to incomplete true library code extraction, \eg an older version of
Admob library contained obfuscated version of code which were not under {\tt
  com.google} domain, thus not counted as true library code.
The poor recall cases were due to excessive true library code inclusion, \eg all
the code of the {\tt Androidify} app was defined under {\tt com.google} domain
which made the distinction of library code obscure.

\subsection{Malware Clustering Results}
\label{subsec:measurement}

\begin{figure}[!h]
    \centering
    \begin{minipage}{0.5\textwidth}
        \centering
        \includegraphics[width=\textwidth, height=4cm]{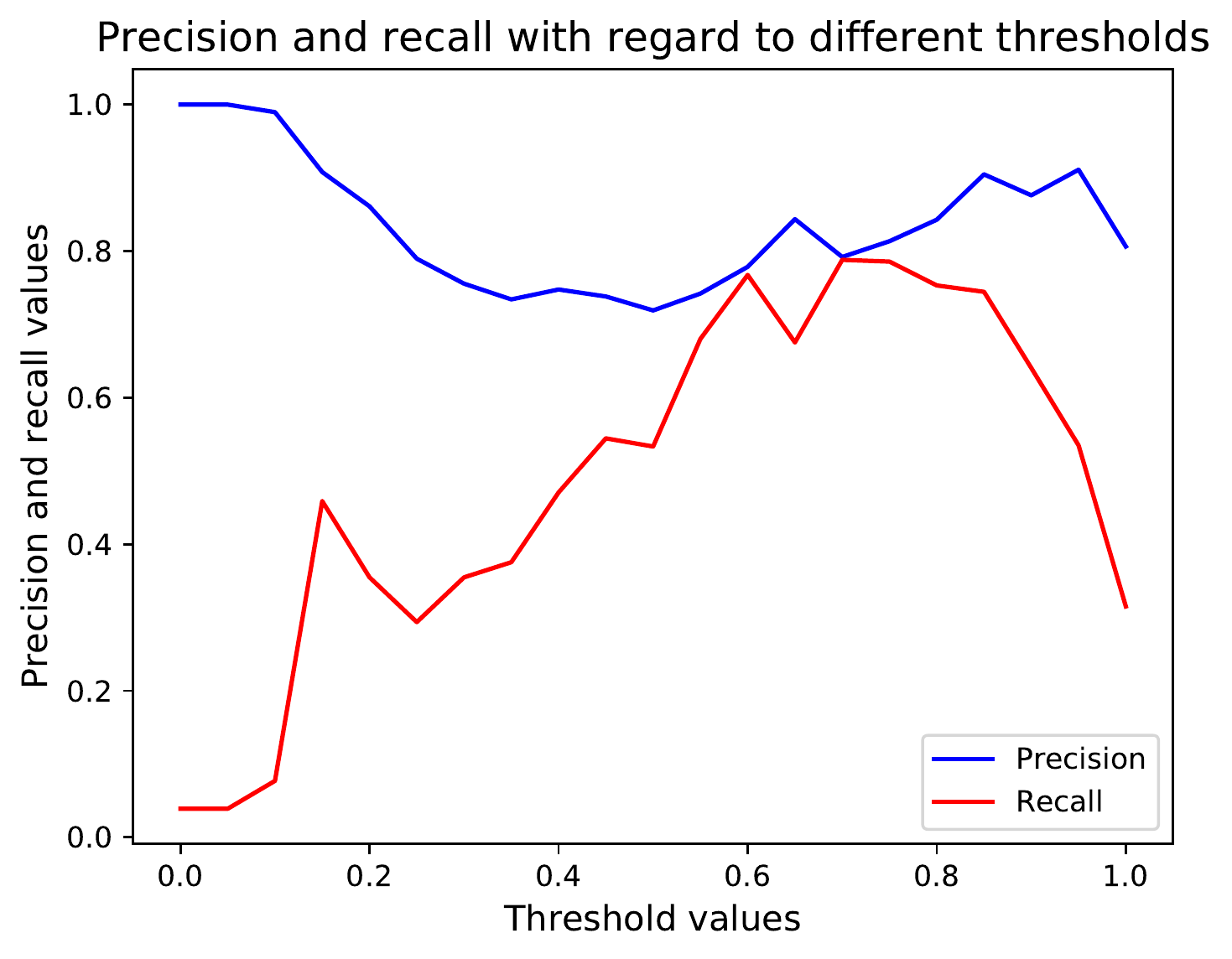}
        \caption{Clustering results of Android Genome malware dataset}
        \label{fig:cluster_genome}
    \end{minipage}%
    \hspace*{0.5cm}
    \begin{minipage}{0.48\textwidth}
        \centering
        \scalebox{0.91}{
          \begin{tabular}{|c||c|c|c|c|}
            \hline
            Datasets & Samples & Clusters & Precision & Recall \\
            \hline
            \hline
            D1 & 1064 & 33 & 0.977 & 0.972 \\
            \hline
            D2 & 1462 &  27 &  0.987 & 0.964 \\
            \hline
            D3 & 1708 &  29 &  0.985 & 0.978 \\
            \hline
            D4 & 1039 & 31 & 0.971 & 0.960 \\
            \hline
            D5 & 2277 & 29 & 0.988 & 0.989 \\
            \hline
            D6 & 1066 & 30 & 0.971 & 0.919 \\
            \hline
            D7 & 1256 & 29 & 0.985 & 0.981 \\
            \hline
            D8 & 1680 & 29 & 0.985 & 0.980  \\
            \hline
            D9 & 2074 & 31 & 0.996 & 0.858 \\
            \hline
            D10 & 1612 & 31 & 0.992 & 0.989 \\
            \hline
          \end{tabular}
        }
        \caption{Clustering results of different sub-version datasets}
        \label{table:multifamily}
    \end{minipage}
\end{figure}

In order to select an optimal clustering threshold for the system and assess the
performance comparing with other known Android malware clustering system, we
first applied our clustering system on the Android Genome malware dataset.
We used the classical precision and recall~\cite{jang2011bitshred,Ye:2010jn,bayer2009scalable,Rieck:2011kw, Hu:2013bg, li2015experimental}
measurements to evaluate the accuracy of clustering results.
Figure~\ref{fig:cluster_genome} describes the clustering precision and recall
results with various thresholds.

The highest F-measure score was 0.82 with precision of 0.90 and
recall of 0.75 when the clustering threshold was 0.85. We set the default
clustering threshold value to be 0.85 for subsequent clustering analysis.
As a reference, ClusTheDroid~\cite{korczynski2015clusthedroid} achieved
precision of 0.74 and recall of 0.73 while clustering
939 of Android Genome malware samples.

Note that the clustering outputs produced by our system is
per sub-version instead of per family, therefore it is more challenging to
properly obtain fine-grained ground truth labels to evaluate the accuracy.
In fact, this was the main reason for a bit low recall of our system with
respect to coarse-grained ground truth labels, \eg
one Android malware family samples might contain multiple versions of malicious
payloads.
While reviewing the clustering results, we noticed that
13 families of the Genome dataset contained more than
one versions of malicious payloads. For example, {\tt Basebridge} contained 7
versions of malicious payloads with threshold of 0.85.

Therefore, we separated the labeled malware samples into sub-versions using the
clustering system,
and further designed several experiments to evaluate the clustering results with
manually verified sub-version ground-truth.
We manually verified the correctness of the sub-version cluster results.
For the generated sub-version clusters, we first checked if the
extracted payload was the indeed malicious.
Since each version of the extracted payloads usually had similar class names and
Dalvik code sequences, the maliciousness of the extracted payload can
be spotted by checking the extracted class names (\eg similar pseudo-random
pattern).
In case the class names were not enough to determine its maliciousness, we then
went through the
reconstructed code segments and checked if there were any suspicious activities
or behaviors, such as stealthily sending out premium SMS.
After verifying the maliciousness of the extracted payload, we then
randomly selected 3 samples from each sub-version group,
and checked if the selected apps contained the same version malicious payload.
Out of 19,725 malware samples that were labeled with 68 families,
we obtained a total of 260 verified sub-version clusters, and each
cluster corresponded to one version of the malicious payloads.

We considered the VirusTotal family labels together with the manually verified
sub-version information as ground truth, and prepared 10 experiment datasets.
For each dataset, we randomly selected 30 sub-versions from the entire
ground truth dataset (\eg 260 sub-versions), then mixed the selected samples together as one input dataset.
The resulting datasets had different overall sizes as each individual sub-version had
different numbers of samples.
The detailed dataset sizes and sample clustering results for the 10 datasets
are presented in Figure~\ref{table:multifamily}.
On average, the sample clustering algorithm separated the input malware samples
into 29.9 clusters, which was extremely close to the reference set (\ie 30
sub-versions).
For the 10 experiment datasets, the clustering algorithm achieved average
precision of 0.984 and average recall of 0.959,
the worst precision and recall for clustering multiple
malware families were 0.971 and 0.858, which suggests that the clustering system
generated consistent and reliable outputs.

\subsection{Key Findings for Malicious Payload Analysis}
\label{subsec:findings}

In this section, we report the key findings learned from the malware sub-version
verification process.

{\bf Significant library code ratio:}
From the labeled malware datasets, we found that the average library code ratio
was larger than 50\% for the malware samples that contained at least one
legitimate library.
This highlights that existing Android malware similarity analysis work becomes
ineffective without properly handling library code.

{\bf Limited versions of malicious payloads:}
During our experiments, we acquired 260 versions of malicious payloads from 68
labeled malware families while conducting clustering of each family.
Among the 68 malware families, 27 families had only one version of malicious
payload, and 5 families had more than 10 different versions of malicious
payloads.
For example, {\tt Dowgin} was the largest malware family and had 23 versions of
malicious payloads extracted.

{\bf Malicious payload under popular namespaces:}
We conducted manual analysis on the extracted malicious payloads, and noted that
29\% of Android malware families
injected their malicious payloads under popular namespaces,
such as {\tt com.google} and {\tt com.android}, or legitimate advertisement
library namespaces like {\tt com.umeng}.
Table~\ref{table:payload} in Appendix includes the detailed malicious payload findings for the identified families.
Since {\tt com.google} and {\tt com.android} are the main class names used by
Android Open Source Project
and Google Mobile Services, such malicious payloads can easily get overlooked.

\section{Limitation}
\label{sec:limitation}

Our Android malware clustering approach is based on the assumption that
malware authors often reuse the same malicious payload to create new malicious
samples, and the obfuscated code sequences of malicious payload would largely remain the
same if they are generated by the same obfuscation tool.
This is consistent with our findings as listed in Section~\ref{subsec:findings}.
Theoretically, advanced obfuscation techniques (e.g., class encryption or
dynamic loading) can eventually break the assumption by generating a new version
of a malicious payload for every new malware instance, or completely removing
the original malicious payload from {\tt classes.dex}.
The attack and defense against malware obfuscation is a long-term arms race, and
has already been observed in the traditional desktop malware analysis
domain.
For example, as observed in desktop malware research~\cite{royal2006polyunpack,
  martignoni2007omniunpack, kang2007renovo}, independent systems might be
desirable to specifically handle the de-obfuscation process.
We consider it as a separate pre-processing step for malware analysis, and leave
a comprehensive solution for advanced obfuscation as an orthogonal problem.
In addition, using dynamic analysis with a sandbox can help further analyze malware. However, dynamic analysis also suffers from its own limitations, such as sandbox evasion and code coverage.

We believe that the Android malware analysis community can benefit from our
work in several aspects.
(a) It offers an alternative malicious payload extraction approach in which we can
extract a more complete version of malicious payloads even if the malicious
payloads are injected under popular library names or under existing functions.
(b) It provides a viable solution for conducting Android malware clustering
analysis
by checking if malware samples contain the same version of malicious payloads.
(c) Majority of Android malware samples are not obfuscated or obfuscated by
simple obfuscation tools, even for the samples we collected recently.
For example, within the extracted 260 versions of malicious payloads, we
observed 181 of them had plain code, and only 79 of them used naming
obfuscation, which was a simple basic obfuscation technique being used in
practice.
(d) As long as there are shared malicious code segments regardless of obfuscation among the samples from the same malware family, our algorithm extracts the shared patterns and uses them for deciding malware clustering output.

\section{Related Work}
\label{sec:related}

\subsection{Android Malware Clustering and App Similarity Analysis}

Due to the challenges that are discussed in Section~\ref{sec:introduction},
existing Android malware clustering approaches have not been widely adopted yet.
ClusTheDroid~\cite{korczynski2015clusthedroid} was a system for clustering
Android malware using 38 features extracted from profiles of reconstructed
dynamic behaviors.
Samra~\cite{samra2013analysis} extracted features from Android app manifest
files, and could only cluster applications into two categories using K-means
algorithm.
Without properly excluding the features or behaviors that belong to the original
benign apps or legitimate libraries, traditional clustering approaches
would not be able to produce promising results.

Similarity analysis is essential for clustering, but existing Android
application similarity analysis techniques were mainly designed to detect
repackaged apps~\cite{zhou2012detecting, hanna2013juxtapp, zhang2014viewdroid},
and such overall similarity analysis based techniques cannot be directly applied for Android malware clustering for reasons described in Section~\ref{sec:introduction}.
SMART~\cite{meng2016semantic} proposed a semantic model for Android malware
analysis, 
but was mainly built for malware detection and classification.
Both Juxtapp~\cite{hanna2013juxtapp} and our system use $n$-gram bytecode features and
feature hashing~\cite{jang2011bitshred,santos2011semi,hu2013mutantx}
as basic building blocks.
However, Juxtapp excluded library code for further analysis if the core application component does not directly invoke it, which still couldn't differentiate a legitimate library and a bogus library with the same legitimate name. Furthermore, directly using Juxtapp to cluster Android malware will suffer the same limitations like other traditional clustering methods as it is based on overall similarity. 

\subsection{Android Malicious Payload Analysis}
Malicious payload identification and extraction is essential for Android malware
analysis.
Zhou and Jiang~\cite{zhou2012dissecting} %
manually analyzed malicious payloads of Android malware and summarized
the findings in the Android Malware Genome project.
DroidAnalytics~\cite{zheng2013droid} presented a multi-level signature based
analytics system to examine and associate repackaged Android malware.
MassVet~\cite{chen2015finding} analyzed graph similarity at the function level and extracted the shared
non-legitimate functions as malicious payloads through commonality and
differential analysis, and it applied a whitelist to exclude legitimate library
code from analysis.

MassVet~\cite{chen2015finding} is close to our work in that both extract
malicious payloads from Android malware.
However, similar to existing Android malware analysis
work~\cite{grace2012riskranker, crussell2012attack, egele2013empirical,
  chen2014achieving, chen2015finding}, MassVet simply used library name based
whitelists to exclude popular library code, which can result in the failure of
malicious payload extraction, and lead to false negatives in malware detection
if malicious payloads are injected under popular library namespaces.
In addition, due to the function level payload granularity of MassVet, it can not be 
easily designed to achieve payload-sharing based Android malware clustering,
since the same function could be shared by different malware families, and the malware 
samples from the same family usually share multiple functions at the same time.
Last but not least, MassVet won't be able to extract malicious payload injected 
under existing functions, while the instruction level payload granularity designed
by our approach enables us to precisely identify one version of malicious payload from each Android malware, which includes all of the malicious components even if they are injected in existing functions or across different functions.

\section{Conclusion}
\label{sec:conclusion}

In this paper, we proposed a practical solution to conduct Android malware
clustering.
As an internal component, the fingerprint based library removal technique
was used to distinguish a legitimate library and a bogus library that may share
the same library name.
Unlike traditional clustering techniques which examine the overall similarity,
we achieved Android malware clustering by checking whether the analyzed Android
malware samples shared the same version of malicious payload code.
Compared with existing malicious payload extraction system, our approach
extracts malicious payloads even if they were injected under popular library
namespaces or under existing benign functions, and it provides a more complete
picture of the whole malicious payload.
Our comprehensive experimental results demonstrate that our clustering approach
generates consistent and reliable outputs with high precision and recall.

\section{Acknowledgment}
\label{sec:ack}

This work was partially supported by the U.S. National Science Foundation under Grant No. 1314925 and 1622402. Any opinions, findings and conclusions or recommendations expressed in this material are those of the authors and do not necessarily reflect the views of the National Science Foundation. 

\bibliographystyle{splncs03}
\bibliography{paper}

\appendix

\section{Detailed malicious payload mining results}

\begin{table}[h!]
\caption{Malicious payload under popular libraries}
\label{table:payload}
\centering
{\scriptsize
\begin{tabular}{|c||c|}
\hline
Family & {Popular Class Names Used}\\
\hline\hline
\multirow{2}{*} {Nickispy} & com.google.android.info.SmsInfo \\
& com.google.android.service.UploadService \\
\hline
\multirow{2}{*} {Uuserv} & com.uuservice.status.SysCaller.callSilentInstall \\
& com.uuservice.status.SilenceTool.MyThread.run \\
\hline
\multirow{2}{*}{Fjcon} & com.android.XWLauncher.CustomShirtcutActivity \\
& com.android.XWLauncher.InstallShortcutReceiver \\
\hline
\multirow{2}{*} {Yzhc} & com.android.Base.Tools.replace\_name \\
& com.android.JawbreakerSuper.Deamon \\
\hline
\multirow{2}{*}{Gumen} & com.umeng.adutils.AdsConnect \\
& com.umeng.adutils.SplashActivity \\
\hline
\multirow{2}{*} {Basebridge} & com.android.sf.dna.Collection \\
& {com.android.battery.a.pa} \\
\hline
\multirow{2}{*} {Spambot} & com.android.providers.message.SMSObserver \\
& com.android.providers.message.Utils.sendSms \\
\hline
\multirow{2}{*} {Moavt} & com.android.MJSrceen.Activity.BigImageActivity \\
& com.android.service.MouaService.InitSms \\
\hline
\multirow{2}{*} {Zitmo} & com.android.security.SecurityService.onStart \\
& com.android.smon.SecurityReceiver.sendSMS \\
\hline
\multirow{2}{*} {Mseg} & com.google.vending.CmdReceiver \\
& android.ad.appoffer.Copy\_2\_of\_DownloadManager \\
\hline
\multirow{2}{*} {Droidkungfu} & com.google.ssearch.SearchService \\
& com.google.update.UpdateService \\
\hline
\multirow{2}{*} {Dowgin} & com.android.qiushui.app.dmc \\
& com.android.game.xiaoqiang.jokes.Data9 \\
\hline
\multirow{2}{*} {Fakeinst} & com.googleapi.cover.Actor \\
& com.android.shine.MainActivity.proglayss\_Click \\
\hline
\multirow{2}{*} {Ksapp} & com.google.ads.analytics.Googleplay\\
& com.google.ads.analytics.ZipDecryptInputStream \\
\hline
\multirow{2}{*} {Bankun} & com.google.game.store.bean.MyConfig.getMsg \\
& com.google.dubest.eight.isAvilible \\
\hline
\multirow{2}{*} {Pjapps} & com.android.MainService.SMSReceiver \\
& com.android.main.TANCActivity \\
\hline
\multirow{2}{*}{Adwo} & com.android.mmreader1030 \\
& com.google.ads.AdRequest.isTestDevice \\
\hline
\multirow{2}{*} {Svpeng} & com.adobe.flashplayer\_.FV.doInBackground \\
& com.adobe.flashplayer\_.FA.startService \\
\hline
\multirow{2}{*} {Opfake} & com.android.appupdate.UpdateService \\
& com.android.system.SurpriseService \\
\hline
\multirow{2}{*} {Badao} & com.google.android.gmses.MyApp \\
& com.android.secphone.FileUtil.clearTxt \\
\hline
\end{tabular}
}
\end{table}

\end{document}